\documentclass[conference]{IEEEtran}
\IEEEoverridecommandlockouts
\usepackage{cite}
\usepackage{amsmath,amssymb,amsfonts}
\usepackage{algorithmic}
\usepackage{graphicx}
\usepackage{textcomp}
\usepackage{xcolor}
\usepackage{booktabs}
\usepackage{multirow}
\usepackage{subfloat}
\usepackage{subcaption}
\def\BibTeX{{\rm B\kern-.05em{\sc i\kern-.025em b}\kern-.08em
    T\kern-.1667em\lower.7ex\hbox{E}\kern-.125emX}}
\begin{document}

\title{ICME 2025 Grand Challenge on Video Super-Resolution for Video Conferencing}
\author{
\IEEEauthorblockN{Babak Naderi, Ross Cutler, Juhee Cho, Nabakumar Khongbantabam, Dejan Ivkovic}
 \IEEEauthorblockA{
        \textit{Microsoft Corporation} \\
         Redmond, USA\\
        \{babaknaderi, ross.cutler, juhcho, nabakumar, dejanivkovic\}@microsoft.com
    }

}

\maketitle

\begin{abstract}
Super-Resolution (SR) is a critical task in computer vision, focusing on reconstructing high-resolution (HR) images from low-resolution (LR) inputs. The field has seen significant progress through various challenges, particularly in single-image SR. Video Super-Resolution (VSR) extends this to the temporal domain, aiming to enhance video quality using methods like local, uni-,  bi-directional propagation, or traditional upscaling followed by restoration.
This challenge addresses VSR for conferencing, where LR videos are encoded with H.265 at fixed QPs. The goal is to upscale videos by a specific factor, providing HR outputs with enhanced perceptual quality under a low-delay scenario using causal models. The challenge included three tracks: general-purpose videos, talking head videos, and screen content videos, with separate datasets provided by the organizers for training, validation, and testing. We open-sourced a new screen content dataset for the SR task in this challenge. Submissions were evaluated through subjective tests using a crowdsourced implementation of the ITU-T Rec P.910. 

\end{abstract}

\begin{IEEEkeywords}
Video super-resolution, Restoration, low-delay
\end{IEEEkeywords}

\section{Introduction}
%Ross or Babak
\label{sec:intro}
Super-Resolution (SR) is a pivotal challenge in the field of computer vision, aiming to reconstruct a high-resolution (HR) image from its low-resolution (LR) counterpart~\cite{son2021ntire}. 
Over the past decade, numerous single image super-resolution challenges have been organized, leading to substantial advancements in the field. These include the Image Super-Resolution~\cite{chen2024ntire, zhang2023ntire, conde2024bsraw, lugmayr2020ntire} and Efficient Super-Resolution~\cite{ren2024ninth, conde2023efficient, zhang2020aim} challenge series.
The Video Super-Resolution (VSR) task extends SR to the temporal domain, aiming to reconstruct a high-resolution video from a low-resolution one. Models for VSR may build upon single image SR techniques, employing various temporal information propagation methods such as local propagation (sliding windows), uni- or bi-directional propagation to enhance quality~\cite{chan2021basicvsr}. Alternatively, traditional upscaling methods like bicubic interpolation can be used, followed by restoration models to improve perceptual quality~\cite{yang2021ntire, valanarasu2023rebotnet}.

VSR has been a focus in challenges such as NTIRE 2019~\cite{nah2019ntire}, NTIRE 2021~\cite{son2021ntire}, and AIM 2024, with the latest exploring efficient VSR~\cite{conde2024aim}. These challenges have addressed various scenarios, including Clean LR~\cite{nah2019ntire, son2021ntire}, LR with motion blur~\cite{nah2019ntire}, and LR with frame drops~\cite{son2021ntire}. The NTIRE 2021 quality enhancement challenge considered input video encoded with H.265 under a fixed quantization parameter (QP) or fixed bitrate~\cite{yang2021ntire} without upscaling. In the AIM 2024 challenge, LR was encoded with AV1 and targeted efficient SR~\cite{conde2024aim}.

In the VSR challenges, the performance of the models is evaluated using objective metrics like PSNR~\cite{gonzalez_digital_2006}, SSIM~\cite{wang_image_2004}, and LPIPS~\cite{zhang_unreasonable_2018}. However, it has been shown that PSNR, SSIM, and MS-SSIM do not correlate well with subjective opinions~\cite{li_toward_2016, seshadrinathan_subjective_2010}, which can lead to misleading model rankings when human users are the target audience. Moreover, models trained on synthetic data often suffer from error propagation when processing videos with various distortions present in real-world recordings~\cite{chan2022investigating}. Some models address this issue by including de-noising as a pre-processing step or limiting the number of frames processed together~\cite{chan2022investigating}. However, our experiments indicate that these approaches can lead to other problems, such as unrealistic videos, flickering, or error propagation in longer sequences ($>$200 frames).
\section{Challenge description}
%  Babak
The ICME grand challenge focuses on video super-resolution for video conferencing, where the low-resolution video is encoded using the H.265 codec with fixed QPs. The goal is to upscale the input LR videos by a specific factor and provide HR videos with perceptually enhanced quality (including compression artifact removal). We follow a \textbf{low-delay scenario} in the entire challenge, where no future frames are used to enhance the current frame. Additionally, there are three tracks specific to video content: \textbf{Track 1}: General-purpose real-world video content (\(\times 4\) upscaling), \textbf{Track 2}: Talking Head videos (\(\times 4\) upscaling), and \textbf{Track 3}: Screen content videos (\(\times 3\) upscaling).

\subsection{Datasets}
Separate training, validation, and test sets were provided for each track. Table~\ref{table:dataset} summarizes the number of clips in each dataset and track. Track 1 used data from the REDS dataset~\cite{Nah_2019_CVPR_Workshops_REDS} for training and validation, and 20 clips from OpenVid-1M~\cite{nan2024openvid} for testing. Track 2 extended the VCD dataset~\cite{naderi2024vcd} with additional real-world video recordings to enhance diversity and realism. Track 3 introduced a new screen content dataset with 95 losslessly recorded desktop videos at 1080p resolution, 30 fps, and durations over one minute. These recordings captured various screen-based scenarios, including applications like Microsoft Word, PowerPoint, Excel, Google Maps, Microsoft Edge, Visual Studio Code, image editing tools, 3D design software, PDF viewers, and multi-window environments, using light or dark themes.

For all tracks, the training set comprised ground truth (GT) videos, where each video clip is 100 frames long, and the downscaled and encoded input videos, which were encoded using the H.265 codec at 6 fixed quantization parameter (QP) levels in the training and validation sets and 4 QPs in the test set to simulate varying compression conditions. Track 1 and Track 2 targeted real-world videos, where other distortions may also be included in the recordings. The video clips in the validation and test sets were 300 frames long to provide a solid basis for subjective testing. The participants were also free to use additional training data.

With this paper, we open-source the training and validation sets for Track 2 (extension of VCD~\cite{naderi2024vcd} for talking head) and Track 3 (screen content)\footnote{The dataset is available at https://github.com/microsoft/VSR-Challenge.}.

\begin{table}[t]
%rewrite
\caption{Clip counts for training, validation, and test sets. }
\label{table:dataset} 
\centering
\huge
\resizebox{\columnwidth}{!}{
    \begin{tabular}{l  c c cc  c c }
    \toprule
     & \multicolumn{2}{c}{\textbf{Training set}} & \multicolumn{2}{c}{\textbf{Validation set}} &\multicolumn{2}{c}{\textbf{Test set}}  \\    
     & \textbf{GT} & \textbf{LR+H.265} & \textbf{GT} & \textbf{LR+H.265} & \textbf{GT} & \textbf{LR+H.265}  \\
    \midrule      
    Track 1 - General & 265 & 1590 &  5 & 30 &  20 & 80  \\
    Track 2 - Talking head & 270 & 1620 & 5 & 30 &  20 & 80 \\
    Track 3 - Screen content & 1520 & 9072&  15 & 90 & 20 & 80  \\

 \bottomrule
    
    \end{tabular}
}
%\vspace{-0.3cm}
\end{table}

\subsection{Evaluation criteria}
\label{sec:eval}
The blind test set for each track was released one week before the challenge deadline. Teams processed the video clips with their models and submitted them for evaluation. Submissions were assessed using the subjective video quality assessment method, specifically the Comparison Category Rating (CCR) test from the crowdsourcing implementation of ITU-T Rec. P.910~\cite{naderi_crowdsourcing_2024}. In CCR, subjects view the source (ground truth) and processed clips, rating the quality of the second clip compared to the first. The presentation order is randomized, and average ratings, presented as CMOS, indicate the processed clip's quality relative to the source. Ratings range from -3 (much worse) to +3 (much better), with 0 indicating no difference. CCR ensures that raters can observe all details in the source video and directly compare the quality of processed clips to source clips.

For the Screen content track, we also included the Character Error Rate (CER) in the challenge score. The CER is determined by applying Optical Character Recognition (OCR) to multiple sections of specific frames in the test set. We annotated at least two frames per video in the test set, with the criterion that the text should be readable in the source video at the target resolution. All annotated areas were successfully detected by OCR and reviewed by an expert. Overall, 210,000 characters were evaluated per submission. We used the normalized average of CMOS and CER, according to Eq.~\ref{eq:score_track3}, as the challenge score for this track.

\begin{equation}
score = \frac{\frac{CMOS}{-3} + CER}{2}
\label{eq:score_track3}
\end{equation}

\subsection{Baseline methods}
%Babak
In addition to traditional upscaling methods such as Bilinear, Bicubic, and Lanczos, we employed two single-image super-resolution models, SwinIR~\cite{liang2021swinir} and Real-SR~\cite{ji2020real}, as well as two video super-resolution models, RealViformer~\cite{zhang2024realviformer} and BasicVSR++~\cite{chan2022basicvsr++}. We utilized these super-resolution models with their pretrained weights as published by their respective authors. As the super-resolution models were trained for \( \times 4 \)  upscaling, we downscaled the resulting videos using Lanczos for the Track~3 test set to achieve 
\( \times 3 \) upscaling. It should be noted that none of the video baseline models are causal and do not follow the challenge criteria, i.e., use future frames when predicting the current frame. We used BasicVSR++ in a way that only one frame is processed at a time to adhere to the challenge rules. However, for RealViformer, we processed in groups of 30 frames \footnote{We observed error propagation when using the default 100-frame package for RealViformer.}.

\section{Challenge Results}
\begin{figure*}
    \centering
    \subfloat[]{\includegraphics[width=0.9\textwidth]{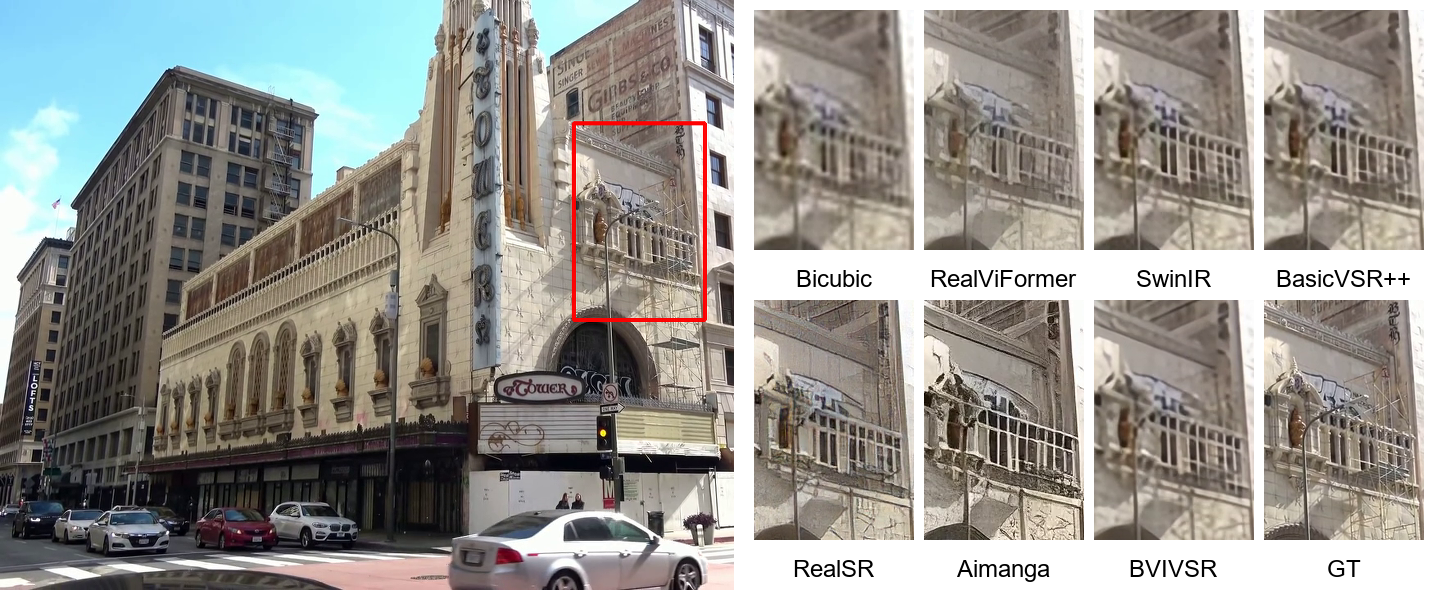}
    \label{fig:track1}
    }

    \subfloat[]{\includegraphics[width=0.9\textwidth]{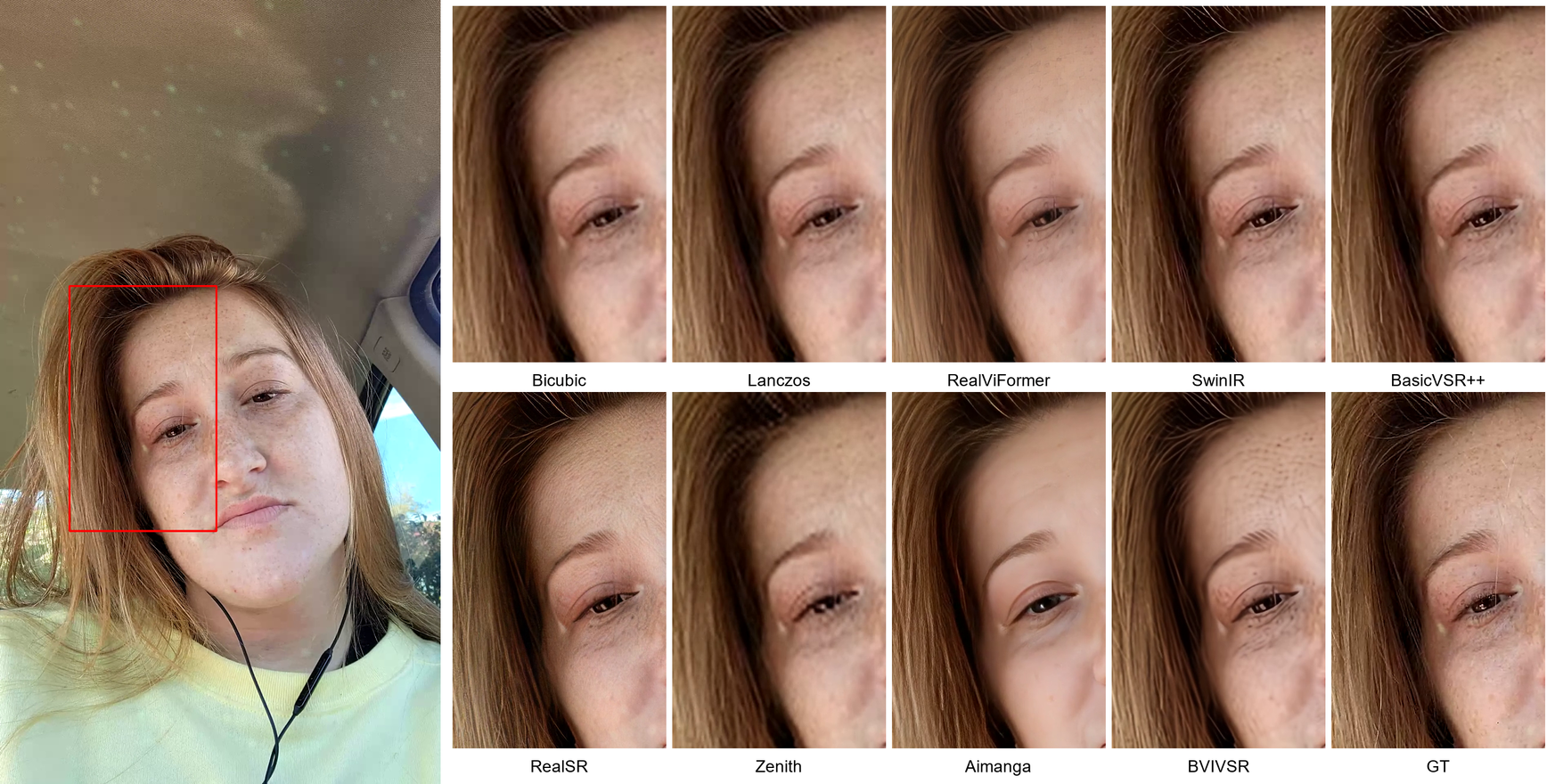} 
    \label{fig:track2}
    }

    \subfloat[]{\includegraphics[width=0.9\textwidth]{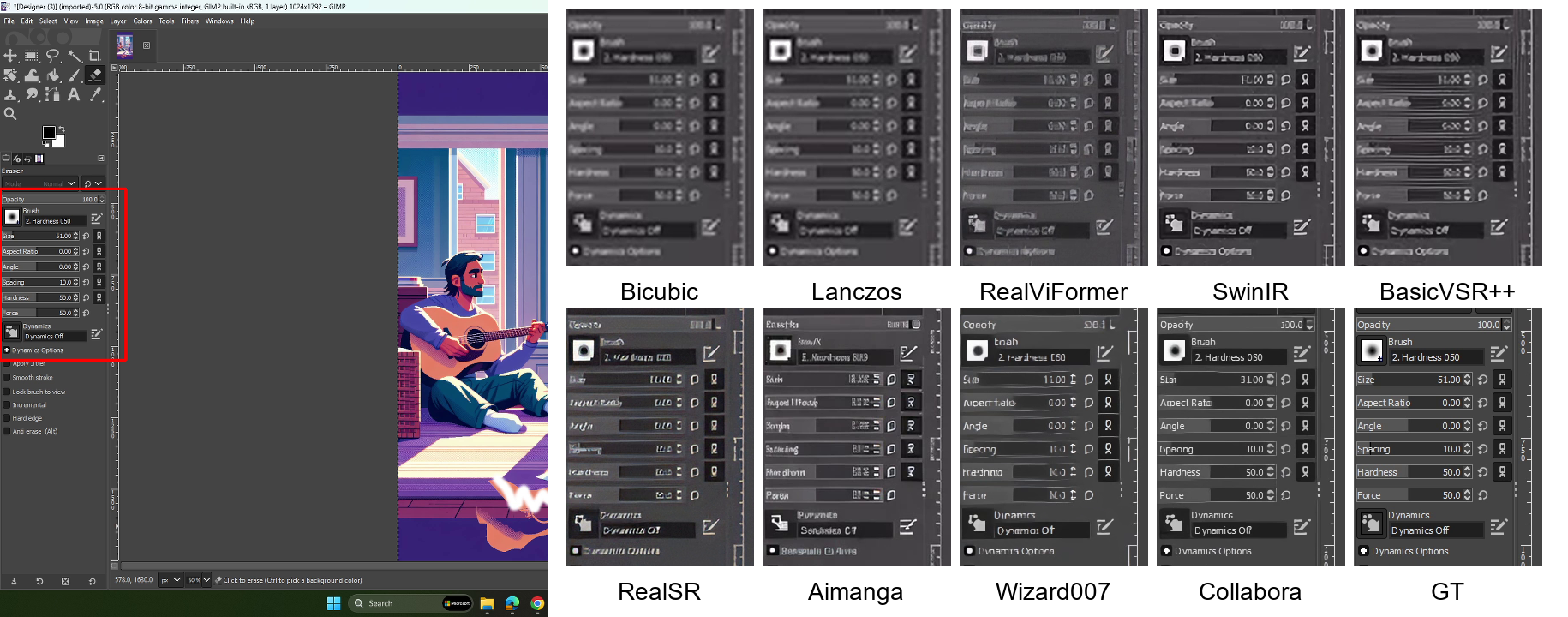}
    \label{fig:track3}
    }

    \caption{Visual comparisons of models evaluated in the challenge for Track 1 (A), Track 2 (B), and Track 3 (C).}
    \label{fig:visual_cmp}
\end{figure*}

%Babak
Overall, five teams participated in the challenge. The average subjective scores in terms of CMOS, multiple objective metrics, and the ranking of models evaluated in the challenge are presented in Table~\ref{table:result}. Two consecutive models are considered to have tied ranks when there is no significant difference between the distributions of CMOS values in Tracks 1 and 2. The ranking in Track 3 is based on the challenge score according to Eq.~\ref{eq:score_track3}. The image-based baseline super-resolution models achieved first place in both the general-purpose and talking-head tracks. However, the participating teams in the screen content track performed significantly better than the baseline models. Visual comparisons of models are presented in~Figure~\ref{fig:visual_cmp}.

\begin{figure}
    \centering
    \subfloat[]{\includegraphics[width=\columnwidth]{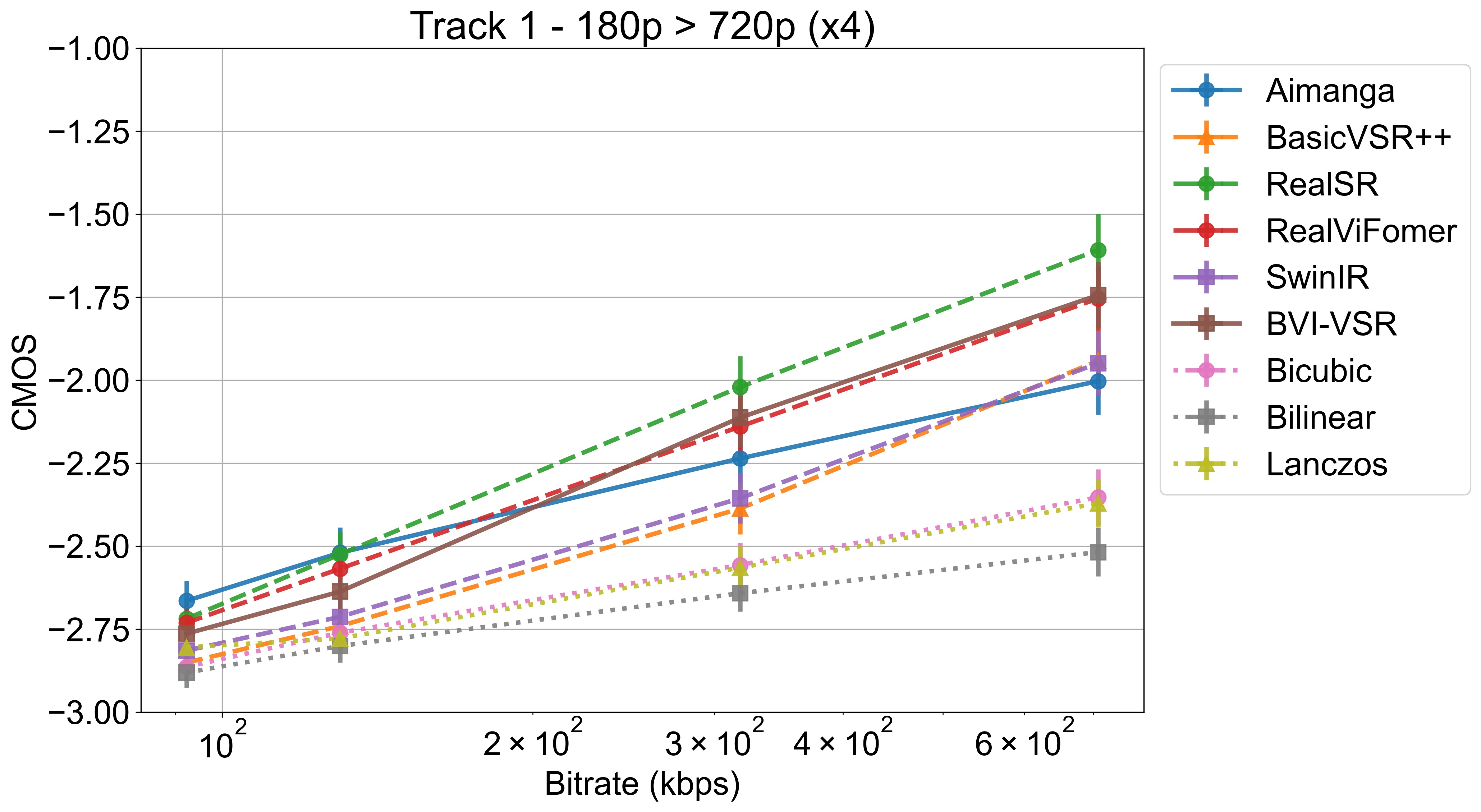}
    \label{fig:cmos_track1}
    }

    \subfloat[]{\includegraphics[width=\columnwidth]{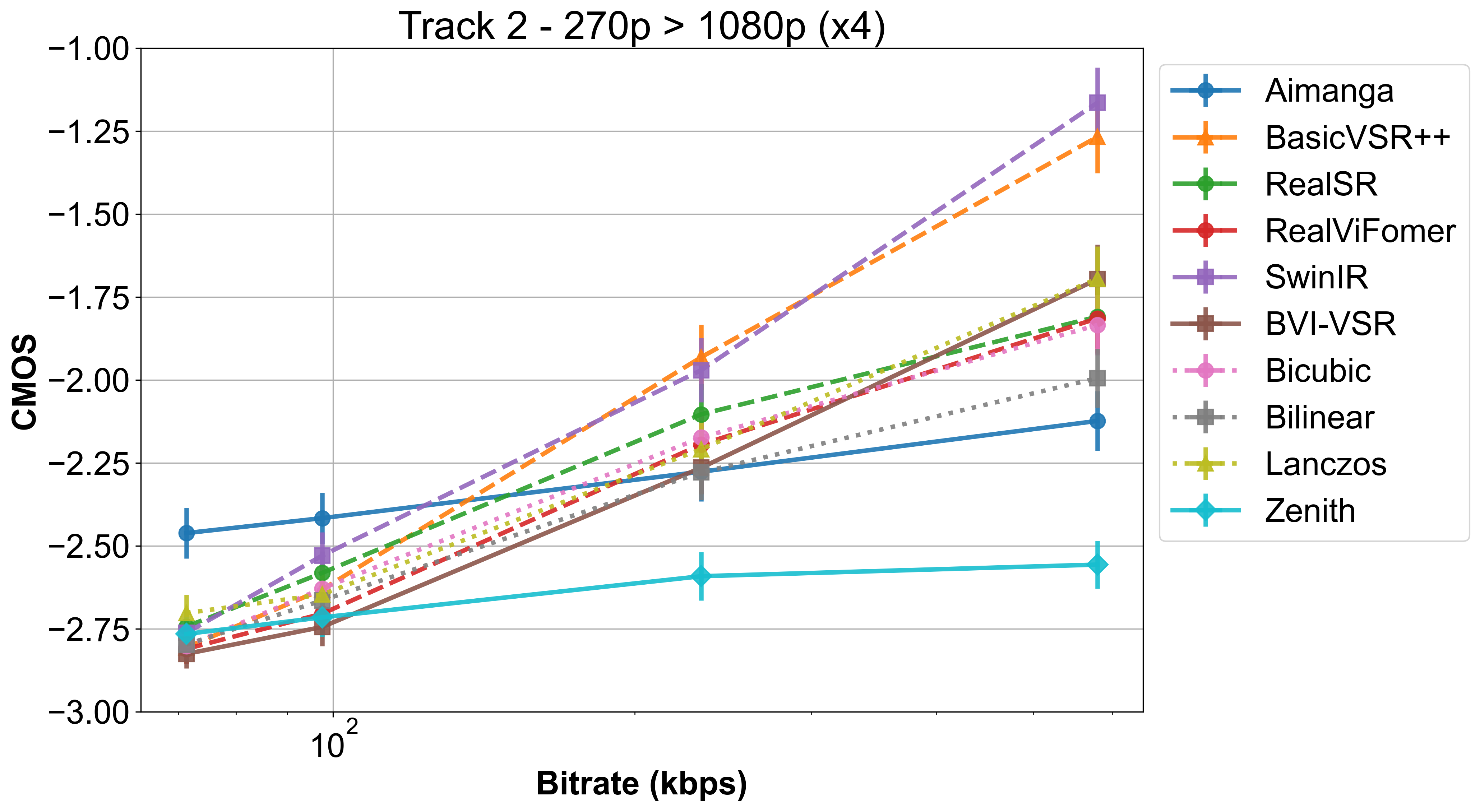}
    \label{fig:cmos_track2}
    }

    \subfloat[]{\includegraphics[width=\columnwidth]{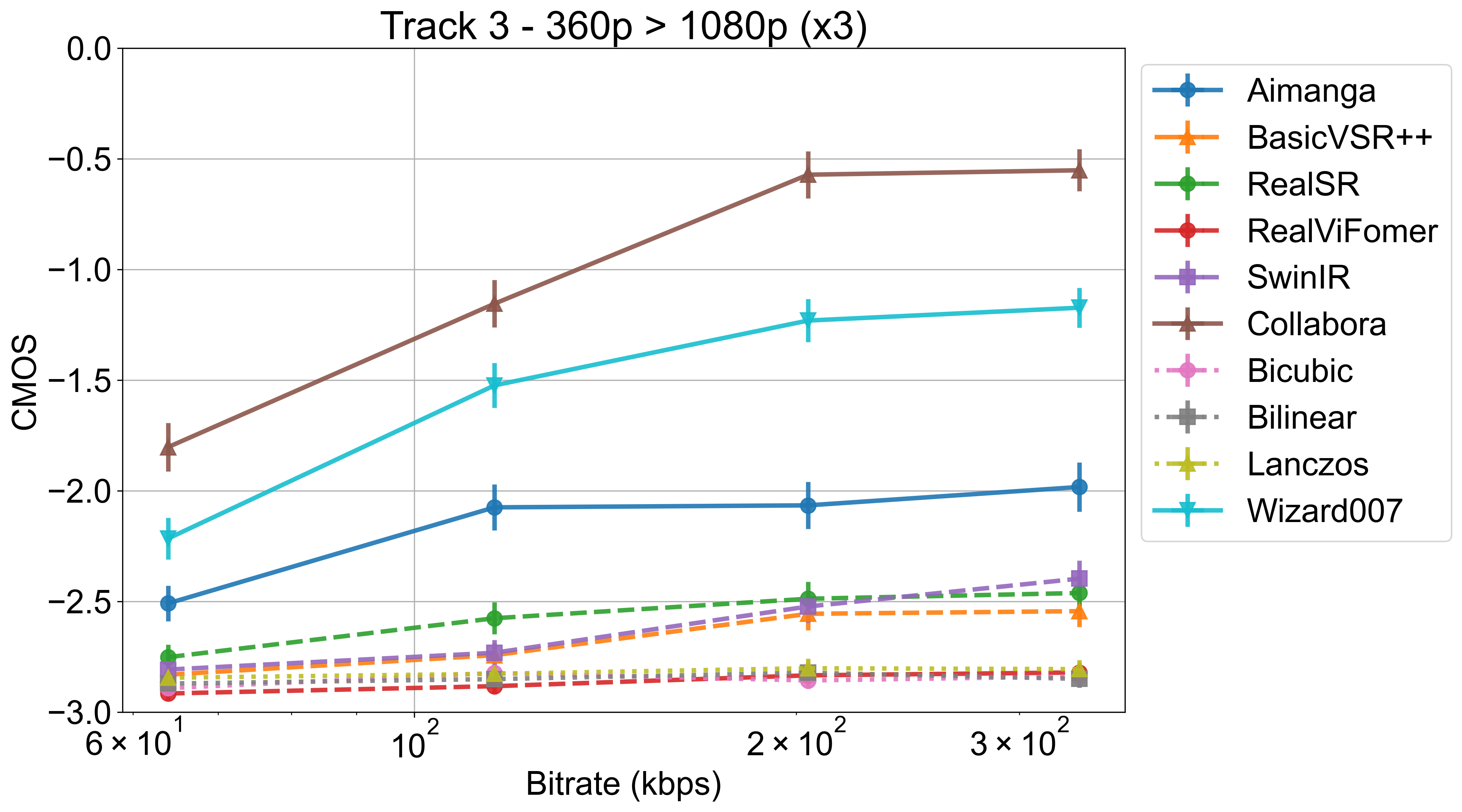}
    \label{fig:cmos_track3}
    }

    \caption{Rate-distortion plots for Track 1 (A), Track 2 (B) and Track 3 (C).}
    \label{fig:rd_plots}
\end{figure}

Figure~\ref{fig:rd_plots} illustrates the rate-distortion plots of the evaluated models. In tracks 1 and 2, a larger difference in performance is observed at higher bitrates where fewer compression artifacts are present. For Track 3, the top-ranked models performed significantly better at all bitrates. Table~\ref{table:cer} presents the CER of each model per quantization parameter. As expected, increasing the compression also increases the CER. We added HEVC at the target resolution to evaluate the performance of the CER measure on a compression artifacts without upscaling. Finally, we applied the CER measure to ground truth (GT) videos to assess test-retest reliability. The results showed a CER of 0.0024, which should be considered the margin of error for this measurement.

\begin{table*}[t]
\caption{Subjective and objective metrics describe the performance of models evaluated in the challenge.}
\label{table:result} 
%\vspace{-0.5cm}
\centering
\resizebox{0.9\textwidth}{!}{
    \begin{tabular}{c l c  c c c c c c }
    \toprule
    \textbf{Track} & \textbf{Method} & \textbf{Rank} &\textbf{CMOS (95\%CI) $\uparrow$} &  \textbf{PSNR $\uparrow$} & \textbf{SSIM $\uparrow$} &  \textbf{VMAF $\uparrow$} & \textbf{LPIPS $\downarrow$} &\textbf{CER $\downarrow$}\\
    \midrule        
    \multirow{8}{*}{Track 1} 
       &RealSR~\cite{ji2020real} & 1 & \textbf{-2.216 (0.047)} & 29.445 & 0.798 & 44.128 & 0.323 & \multirow{8}{*}{--}  \\
&RealViFormer~\cite{zhang2024realviformer} & 2 & -2.295 (0.045) & 29.741 & 0.804 & 28.669 & \textbf{0.274} &  \\
&BVIVSR & 2 & -2.316 (0.044) & \textbf{31.027} & \textbf{0.846} & \textbf{53.415} & 0.316 &  \\
%they used InvSR for track1 and 2
&Aimanga & 2 & -2.355 (0.044) & 26.388 & 0.752 & 51.086 & 0.312 &  \\
&SwinIR~\cite{liang2021swinir} & 3 & -2.459 (0.040) & 30.266 & 0.822 & 47.798 & 0.346 &  \\
&BasicVSR++~\cite{chan2022basicvsr++} & 3 & -2.481 (0.039) & 30.289 & 0.821 & 47.714 & 0.333 &  \\
&Lanczos & 4 & -2.631 (0.032) & 30.345 & 0.817 & 34.507 & 0.443 &  \\
&Bicubic & 4 & -2.633 (0.033) & 30.262 & 0.814 & 31.605 & 0.444 &  \\
&Bilinear & 5 & -2.712 (0.029) & 29.937 & 0.803 & 19.543 & 0.454 &  \\

    \midrule
    \multirow{10}{*}{Track 2} 
&SwinIR~\cite{liang2021swinir} & 1 & \textbf{-2.122 (0.053)} & 35.308 & 0.929 & 55.262 & 0.231 & \multirow{10}{*}{--} \\
&BasicVSR++~\cite{chan2022basicvsr++} & 2 & -2.189 (0.052) & 35.385 & 0.930 & 54.956 &\textbf{0.224} &  \\
&Aimanga & 3 & -2.322 (0.042) & 30.808 & 0.882 & 58.313 & 0.273 &  \\
&RealSR~\cite{ji2020real} & 3 & -2.332 (0.043) & 34.853 & 0.917 & 53.775 & 0.291 &  \\
&Lanczos & 3 & -2.341 (0.043) & 35.957 & 0.929 & 43.304 & 0.276 &  \\
&Bicubic & 3 & -2.364 (0.042) & 35.869 & 0.929 & 40.688 & 0.279 &  \\
&BVIVSR & 4 & -2.427 (0.043) & \textbf{35.986} & \textbf{0.939} & \textbf{61.661} & 0.227 &  \\
&RealViFormer~\cite{zhang2024realviformer} & 4 & -2.427 (0.042) & 34.724 & 0.923 & 43.153 & 0.225 &  \\
&Bilinear & 4 & -2.429 (0.040) & 35.507 & 0.927 & 29.395 & 0.289 &  \\
&Zenith & 5 & -2.661 (0.032) & 33.643 & 0.915 & 35.127 & 0.287 &  \\

    \midrule
    \multirow{9}{*}{Track 3} 
    % slightly different than what piblished in ICME as CER claculation is impro
& Collabora & 1 & \textbf{-1.029 (0.059)} & \textbf{34.301} & \textbf{0.953} & \textbf{84.143} &\textbf{0.086} & \textbf{0.139} \\
 & Wizard007 & 2 & -1.554 (0.053) & 32.397 & 0.939 & 78.513 & 0.110 & 0.215 \\
 & Aimanga & 3 & -2.155 (0.052) & 27.642 & 0.890 & 48.889 & 0.153 & 0.625 \\
 & SwinIR~\cite{liang2021swinir} & 4 & -2.617 (0.034) & 29.575 & 0.912 & 74.085 & 0.158 & 0.566 \\
 & Lanczos & 5 & -2.820 (0.020) & 28.579 & 0.883 & 52.922 & 0.332 & 0.545 \\
 & Bicubic & 6 & -2.853 (0.019) & 28.493 & 0.882 & 50.367 & 0.333 & 0.555 \\
 & BasicVSR++~\cite{chan2022basicvsr++} & 7 & -2.669 (0.032) & 29.499 & 0.909 & 70.175 & 0.183 & 0.647 \\
 & Bilinear & 8 & -2.848 (0.019) & 28.209 & 0.877 & 39.237 & 0.321 & 0.606 \\
 & RealSR~\cite{ji2020real} & 9 & -2.570 (0.036) & 28.647 & 0.896 & 60.819 & 0.152 & 0.703 \\
 & RealViFormer~\cite{zhang2024realviformer} & 10 & -2.864 (0.018) & 27.928 & 0.886 & 25.881 & 0.220 & 0.903 \\
 
    \bottomrule
    
    \end{tabular}
}
%\vspace{-0.3cm}
\end{table*}

\begin{table}[t]
%rewrite
\caption{CER measured for Track3, organized per Quantization Parameters (QP).}
\label{table:cer} 
\centering
\resizebox{\columnwidth}{!}{
    \begin{tabular}{l c  c c c c c }
    \toprule
    \textbf{Method} & \textbf{QP=17} & \textbf{QP=22} &\textbf{QP=27} &  \textbf{QP=32} & \textbf{Overall} \\
    \midrule        
HEVC@1080p & 0.004 & 0.004 & 0.004 & 0.005 & 0.004 \\
Collabora & 0.063 & 0.075 & 0.123 & 0.297 & 0.139 \\
Wizard007 & 0.132 & 0.151 & 0.207 & 0.371 & 0.215 \\
Lanczos & 0.482 & 0.493 & 0.548 & 0.657 & 0.545 \\
Bicubic & 0.492 & 0.507 & 0.557 & 0.662 & 0.555 \\
SwinIR~\cite{liang2021swinir} & 0.451 & 0.487 & 0.585 & 0.739 & 0.566 \\
Bilinear & 0.553 & 0.568 & 0.608 & 0.695 & 0.606 \\
Aimanga & 0.582 & 0.595 & 0.625 & 0.697 & 0.625 \\
BasicVSR++~\cite{chan2022basicvsr++} & 0.571 & 0.584 & 0.662 & 0.771 & 0.647 \\
RealSR~\cite{ji2020real} & 0.662 & 0.677 & 0.701 & 0.772 & 0.703 \\
RealViFormer~\cite{zhang2024realviformer} & 0.877 & 0.887 & 0.911 & 0.937 & 0.903 \\

 \bottomrule    
\end{tabular}
}
%\vspace{-0.3cm}
\end{table}

\subsection{Comparison of subjective and objective metrics}
% Babak
% Include PCC for PSNR, SSIM, VMAF, LPIPS with P.910
We compare the correlation coefficients between subjective and objective metrics using Pearson, Spearman, Kendall’s Tau-b, and a variant of Tau-b that considers the distribution of ratings and objective metrics (hereafter referred to as Tau-b95). To do so, we first create a rank order of the items based on the CMOS and the 95\% confidence interval (CI) values from the subjective test, following the method described in \cite{naderi2020transformation}. Specifically, two items are considered to have tied ranks when the CMOS value of one item falls within the 95\% CI of the other.

Subsequently, we compute Kendall’s Tau-b on the resulting rank-ordered list (referred to as Tau-b95). For objective metrics, we use the distribution of metric values over individual clips and calculate the 95\% CI accordingly.

The correlation coefficients between subjective and objective metrics are presented in Table~\ref{table:subj_obj} and Table~\ref{table:subj_obj_taub}. For Tracks 1 and 2, Pearson and Spearman correlations between subjective quality scores and both PSNR and SSIM are very weak, while VMAF shows a moderate correlation. LPIPS demonstrates a strong Pearson correlation with subjective ratings; however, its Spearman correlation coefficient of $\leq 0.8$ highlights the necessity of using subjective quality evaluations for ranking video super-resolution models. Similar trends are observed with the Tau-b and Tau-b95 coefficients.

In contrast, for screen content, PSNR and SSIM exhibit strong Pearson and Spearman correlations with subjective ratings, and CER, indicating their suitability for use during model development in this context, particularly given the importance of preserving text content. For this dataset, LPIPS shows the highest rank-based correlation with subjective scores; however, a Tau-b coefficient of $\leq 0.84$ still underscores the need for subjective testing when ranking models for track3.

\begin{table}[t]
%rewrite
\caption{Pearson (upper triangular) and Spearman (lower triangular) correlation coefficients between subjective and objective metrics in the model level. }
\label{table:subj_obj} 
\centering
\resizebox{1\columnwidth}{!}{
    \begin{tabular}{c l  c c c c c c }
    \toprule
    \textbf{Track} & \textbf{Metric} & \textbf{CMOS} &\textbf{PSNR} &  \textbf{SSIM} &\textbf{VMAF} & \textbf{LPIPS} & \textbf{CER} \\
    \midrule        
 
\multirow{5}{*}{Track 1}  
 & CMOS &  & -0.212 & -0.073 & 0.581 &\textbf{ -0.893} &\multirow{5}{*}{--}   \\
 & PSNR & -0.164 &  & 0.958 & -0.098 & 0.165 &  \\
 & SSIM & 0.027 & 0.897 &  & 0.161 & -0.012 &  \\
 & VMAF & 0.391 & 0.318 & 0.519 &  & -0.641 &  \\
 & LPIPS & \textbf{-0.800} & 0.127 & -0.096 & -0.464 &  &  \\

  \midrule  
 
\multirow{5}{*}{Track 2}  
 & CMOS &  & 0.232 & 0.316 & 0.598 & \textbf{-0.732} &  \multirow{5}{*}{--} \\
 & PSNR & 0.077 &  & 0.914 & -0.157 & -0.323 &  \\
 & SSIM & 0.443 & 0.648 &  & 0.047 & -0.468 &  \\
 & VMAF & 0.430 & -0.135 & 0.429 &  & -0.516 &  \\
 & LPIPS & \textbf{-0.720} & -0.248 & -0.613 & -0.328 &  &  \\

  \midrule  

\multirow{6}{*}{Track 3}   
 &  CMOS &  & 0.885 & \textbf{0.894} & 0.725 & -0.783 & -0.813 \\
 & PSNR & 0.821 &  & 0.957 & 0.841 & -0.714 & -0.893 \\
 & SSIM & 0.902 & 0.941 &  & 0.879 & -0.861 & -0.847 \\
 & VMAF & 0.814 & 0.946 & 0.923 &  & -0.722 & -0.811 \\
 & LPIPS & \textbf{-0.929} & -0.818 & -0.923 & -0.782 &  & 0.583 \\
 & CER & -0.779 & -0.861 & -0.775 & -0.786 & 0.682 &  \\

 \bottomrule    
\end{tabular}
}
%\vspace{-0.3cm}
\end{table}

\begin{table}[t]
%rewrite
\caption{Kendall’s Tau-b (upper triangular) and Tau-b95 (lower triangular) correlation coefficients between subjective and objective metrics at the model level. }
\label{table:subj_obj_taub} 
\centering
\resizebox{1\columnwidth}{!}{
    \begin{tabular}{c l  c c c c c c }
    \toprule
    \textbf{Track} & \textbf{Metric} & \textbf{CMOS} &\textbf{PSNR} &  \textbf{SSIM} &\textbf{VMAF} & \textbf{LPIPS} & \textbf{CER} \\
    \midrule        
 
\multirow{5}{*}{Track 1}  

& CMOS &  & -0.091 & 0.127 & 0.382 & \textbf{-0.673} &\multirow{5}{*}{--}\\
& PSNR & -0.025 &  & 0.782 & 0.236 & 0.055 & \\
& SSIM & 0.201 & 0.788 &  & 0.455 & -0.091 & \\
& VMAF & 0.426 & 0.104 & 0.466 &  & -0.345 & \\
& LPIPS & \textbf{-0.692} & -0.136 & -0.317 & -0.466 & & \\

  \midrule   
  \multirow{5}{*}{Track 2}  
 & CMOS &  & 0.096 & 0.362 & 0.288 & \textbf{-0.583} &\multirow{5}{*}{--}\\
& PSNR & 0.196 &  & 0.529 & -0.103 & -0.235 & \\
& SSIM & 0.236 & 0.406 &  & 0.309 & -0.471 & \\
& VMAF & 0.357 & -0.163 & 0.230 &  & -0.250 & \\
& LPIPS & \textbf{-0.556} & -0.203 & -0.380 & -0.208 &  &\\

  \midrule  
  \multirow{5}{*}{Track 3}  
& CMOS &  & 0.752 & 0.752 & 0.676 & \textbf{-0.790} & -0.638 \\
& PSNR & 0.718 &  & 0.848 & 0.848 & -0.657 & -0.733 \\
& SSIM & 0.744 & 0.888 &  & 0.771 & -0.810 & -0.657 \\
& VMAF & 0.693 & 0.801 & 0.845 &  & -0.581 & -0.695 \\
& LPIPS &\textbf{ -0.839} & -0.685 & -0.776 & -0.661 &  & 0.543 \\
& CER & -0.720 & -0.746 & -0.772 & -0.668 & 0.630 &  \\

 \bottomrule    
\end{tabular}
}
%\vspace{-0.3cm}
\end{table}

\section{Challenge Methods and Teams}
% each reviwer should give a summary of paper
% provide Description and Implementation details for each paper
Below, the model for each participating team is briefly described. The complexity of the models is reported in Table~\ref{table:complexity}.

\begin{table*}[t]
\caption{Comparison of method complexity for challenge participating teams and baseline models.}
\label{table:complexity} 
%\vspace{-0.5cm}
\huge
\centering
\resizebox{0.9 \textwidth}{!}{
    \begin{tabular}{ l c c c c c c c }
    \toprule
    \textbf{Method} & \textbf{Param (M)} &\textbf{FLOPs (G)} &\textbf{Track} & \textbf{Input size} & \textbf{Latency (s)} & \textbf{Throughput (FPS)}  &\textbf{Hardware} \\
   
    \midrule        
    Aimanga    &  16.7 &  73.5 & Track 2 &  [1, 3, 64, 64]  & 3.12  &  0.32  & GeForce RTX 4090  \\
    \midrule
    \multirow{2}{*}{BVIVSR} & \multirow{2}{*}{5.3} & 455.16 & Track 1 & [1, 3, 180, 320] & 0.141 & 7.11 &  \multirow{2}{*}{GeForce RTX 4090}\\
    & & 455.16 & Track 2 & [1, 3, 270, 480] & 0.375 & 2.67 &  \\
     \midrule 
     Collabora & 983   & 6,404  & Track 3 & [1, 3, 360, 640] & 23.8  & 0.042   & GeForce RTX 4090 \\
     \midrule  
     Wizard007 & 22.6  &  10,645 & Track 3 & [1, 3, 360, 640] &  2.49 & 0.402 &  Tesla V100-32GB \\

     \midrule  
     Zenith     &  1.9  & 50.61 & Track 2 & [8, 5, 3, 64, 64] & 0.049  & 30    &  GeForce RTX 3090 \\ 
    \midrule 

      \multirow{3}{*}{BasicVSR++}& \multirow{3}{*}{7.3} & 280.5  & Track 1 & [1, 3, 180, 320] & 0.050 & 19.73  & \multirow{3}{*}{Tesla V100-32GB} \\
 & & 635.8 & Track 2 & [1, 3, 270, 480] & 0.104  & 9.51  &   \\
 & & 1,122 & Track 3 & [1, 3, 360, 640] & 0.187  & 5.32  &   \\
\midrule 

 \multirow{3}{*}{Real-SR}& \multirow{3}{*}{16.7} & 1034 & Track 1 & [1, 3, 180, 320] & 0.153 & 6.53  & \multirow{3}{*}{Tesla V100-16GB} \\
 & & 2343 & Track 2 & [1, 3, 270, 480] & 0.338  & 2.95  &   \\
 & & 4135 & Track 3 & [1, 3, 360, 640] & 0.599  & 1.66  &   \\

\midrule 

     \multirow{3}{*}{RealViformer}& \multirow{3}{*}{5.8} & 123.5 & Track 1 & [1, 3, 180, 320] &   0.050  &  19.94 & \multirow{3}{*}{Tesla V100-16GB}  \\
     & & 280.0 & Track 2 & [1, 3, 270, 480] & 0.085  &  11.81 & \\
     & & 494.2 & Track 3 &[1, 3, 360, 640] &  0.143  &  6.98 &  \\

\midrule 

\multirow{3}{*}{SwinIR}& \multirow{3}{*}{0.91} & 176.98 & Track 1 & [1, 3, 180, 320] &  0.445 & 2.21  & \multirow{3}{*}{Tesla V100-16GB} \\
 & & 841.6 & Track 2 & [1, 3, 270, 480] &  1.086  & 0.91  &   \\
 & & 2,589 & Track 3 & [1, 3, 360, 640] &  1.949  & 0.52 &   \\

    \bottomrule    
    \end{tabular}
}
%\vspace{-0.3cm}
\end{table*}

\subsection{Team Aimanga}
Aimanga team's model for Track 2 is based on the RealESRGAN architecture\footnote{For Track 1 and 3, Aimanga team used InvSR~\cite{yue2024arbitrary} with no or minor changes.}. The generator includes a feature extraction convolution layer, 23 Residual-in-Residual Dense Blocks (RRDBs), and PixelShuffle up-sampling for reconstruction. The discriminator uses a U-Net architecture, and the model processes one frame at a time. The CodeFormer model is used for post-processing to enhance visual quality. They propose sub-sampling videos into tiles for training, selecting those with the largest gradients using a Laplacian operator. They also explore various post-processing effects on these tiles. The model was trained on the nomos2k dataset, with preprocessing to select tiles and augmentation through various degradations (blur, noise, compression artifacts). Visual comparisons in Figure~\ref{fig:track2} show that this model significantly removes details from faces

\subsection{Team BVIVSR}
%Ross
Team BVIVSR introduced the VSR-HE model, built on a Hierarchical Encoding Transformer (HiET) architecture, using multiple transformer blocks arranged in a pyramid to capture local details and long-range context. The network is trained in two stages: first with a combined pixel-level and structural-similarity loss (for fidelity), then fine-tuned with a GAN-based adversarial loss to enhance sharpness and realism. Besides the training set from the challenge, the authors used the BVI-AOM video dataset, augmented using the same approach, for training. They also reported significant gains in objective metrics (PSNR, SSIM, MS-SSIM, VMAF), which are in line with the challenge results.

\subsection{Team Collabora}
%Dejan
The Collabora team propose an innovative method of combining several existing, state-of-the-art super-resolution and text recovery methods to enhance the quality of screen content image (SCI). They use Implicit Transformer Super-Resolution Network (ITSRN) to upscale low-resolution SCI and then OCR to detect text regions. They correct the text within those regions by using Large Language Model (LLM) to infer correct text and then re-synthesize the text regions using TextSSR which uses LLM's text output and ITSRN's image output. 
A new dataset, Screen Content Dataset (SCD), was created and used to train TextSSR. It includes diverse screen content scenarios and detailed text region annotations. Visual comparisons in Figure~\ref{fig:track3} show impressive results, but significant latency makes this approach unsuitable for real-time communication. 

\subsection{Team Wizard 007}
%Juhee
The Wizard 007 team propose a novel method for screen content track built upon a pretrained ITSRN, chosen for its ability to model continuous image representations and capture long-range dependencies—critical for preserving fine-grained textual and structural details in screen content. To further tailor the model for screen content scenarios, they introduce a composite loss function that integrates perceptual quality, VGG-based semantic similarity, and a CER loss. This combination ensures that the super-resolved outputs are not only visually appealing but also semantically accurate, especially in text-heavy regions. Their model is fine-tuned specifically on text-rich patches—defined as those with over 20\% textual content—identified using EasyOCR, ensuring that the training process emphasizes readability and functional clarity. Although they achieved 2nd rank, the latency of this model is significantly lower than that of the winning team.

\subsection{Team Zenith}
Team Zenith's VSR model integrates spatial and temporal modeling through four stages: local feature extraction, temporal modeling, reconstruction, and region-specific enhancement. It employs convolutional encoders, residual blocks with channel attention, attention-based recurrent units, subpixel reshaping, and a pretrained locator for refining key areas like faces. The network was trained using an extended VCD dataset and a composite loss function, which includes reconstruction loss, edge-preserving loss for sharpness in high-frequency regions, and perceptual loss for semantic consistency. These components are weighted and combined to balance detail restoration with temporal stability. Figure~\ref{fig:track2} illustrates that this model adds artifacts specifically in edge areas.

\section{Conclusion}
We organized the Video Super-Resolution for Video Conferencing challenge in three content-based tracks. In general-purpose and talking head tracks, baseline models outperformed participating teams in subjective quality tests. However, in the screen content track, participating teams significantly outperformed the baselines, utilizing OCR and CER. We also found a low correlation between objective metrics and subjective quality scores in general-purpose and talking head contexts, highlighting the necessity of subjective evaluation for ranking model performance in these areas. Additionally, we introduced a new open-sourced screen content dataset and extended VCD dataset for SR and other relevant tasks.

\bibliographystyle{IEEEbib}
\bibliography{IC3-AI,main,sr}

\end{document}